\documentclass[%
 reprint,
superscriptaddress,
 amsmath,amssymb,
 aps,
prb, citeautoscript,
]{revtex4-2}

\usepackage{graphicx}
\usepackage{dcolumn}
\usepackage{bm}
\usepackage{color, soul}
\usepackage{natbib}
\usepackage[colorlinks, linkcolor=blue, citecolor=blue, urlcolor=blue]{hyper ref}
\makeatletter
\newcommand*{\rom}[1]{\expandafter\@slowromancap\romannumeral #1@}
\makeatother

\begin{document}

\preprint{APS/123-QED}

\title{Surface phase diagram of CsSnI$_3$ from first-principles calculations}

\author{Kejia Li}
\affiliation{Department of Electrical and Computer Engineering, State University of New York at Binghamton, Binghamton, New York 13902, United States}
\affiliation{Materials Science and Engineering Program, State University of New York at Binghamton, Binghamton, New York 13902, United States}
\author{Chadawan Khamdang}
\affiliation{Department of Electrical and Computer Engineering, State University of New York at Binghamton, Binghamton, New York 13902, United States}
\author{Mengen Wang}
\email{mengenwang@binghamton.edu}
\affiliation{Department of Electrical and Computer Engineering, State University of New York at Binghamton, Binghamton, New York 13902, United States}
\affiliation{Materials Science and Engineering Program, State University of New York at Binghamton, Binghamton, New York 13902, United States}

\begin{abstract}
CsSnI$_3$ is widely studied as an environmentally friendly Pb-free perovskite material for optoelectronic device applications. 
To further improve material and device performance, it is important to understand the surface structures of CsSnI$_3$.
We generate surface structures with various stoichiometries, perform density functional theory calculations to create phase diagrams of the CsSnI$_3$ (001), (110), and (100) surfaces, and determine the most stable surfaces under a wide range of Cs, Sn, and I chemical potentials.
Under I-rich conditions, surfaces with Cs vacancies are stable, which lead to partially occupied surface states above the valence band maximum.
Under I-poor conditions, we find the stoichiometric (100) surface to be stable under a wide region of the phase diagram, which does not have any surface states and can contribute to long charge carrier lifetimes.
Consequently, the I-poor (Sn-rich) conditions will be more beneficial to improve the device performance.
\end{abstract}

\maketitle

\section{\label{intro}Introduction}

Lead-based halide perovskites are widely studied as a light-absorber for high-efficiency solar cells due to the large absorption coefficients, low effective mass, and long carrier lifetimes \cite{jeong2020_pb_perovskite_25,kim2020pervoskite_review, chen2016carrier}.
However, Pb is a toxic material, which motivate the study of Pb-free perovskites by replacing Pb with less toxic divalent cations including Sn, Ge, and Cu \cite{ke2019prospects_pbfree,wang2021leadfree}.

Sn-based perovskites have been proposed as the most promising lead-free perovskite.
The all-inorganic CsSnI$_3$ with a bandgap of 1.3 eV and a promising theoretical power conversion efficiency (PCE) limit of 33\% is a widely investigated Sn-based perovskite material \cite{yu2011cssni3_bandgap,ruhle2016ShockleyQueisser_limit}.
The experimentally achieved PCE of CsSnI$_3$-based device, however, is well below the theoretical limit \cite{zhou2022sn_perovskite_pce14, yu2021sn_perovskite_pce14,hossain2023cssni3_24}.
Experimental and theoretical studies have been devoted to revealing the origin of low PCE of CsSnI$_3$.
The self $p$-type doping and low material stability of CsSnI$_3$ have been proposed to be the main factors that hinder the performance of CsSnI$_3$ \cite{zhang2022cssni3_defect_dft,hossain2023cssni3_24}.

Defects on the surface and grain boundaries of perovskites may have different defect properties compared with defects in the bulk, which can possibly lead to deep trap levels and become non-radiative recombination centers \cite{park2019grain_boundary_defects_perovkites}.
To understand the role of surface and surface defects in the device performance, it is important to understand the surface structures.
There is a growing interest in density functional theory (DFT) calculations of Pb-based perovskite surfaces properties including surface energies of MAPbI$_3$, CsPbI$_3$, CsPbBr$_3$, and CsPbCl$_3$  \cite{yang2021cubic_cspbbr3_surface, yang2022surface_ortho_cspbx3, haruyama2016mapbi_surface,haruyama2014mapbi3_surface}, defect formation on perovskite surfaces \cite{liu2017mapbbr3_surface_defect, ten2019mapbbr3_surface_defect,han2022cspbi3_surface_defects}, and interactions between passivating agents with the perovskite surfaces \cite{smart2021ligand_cspbbr3,yoo2020ligand_cspbbr3_surface, he2019cspbi3_ms2,zheng2020cssni2_ligand_passivate}.

For Sn-based perovskites, surface properties also play an important role in the material instability that limits the device performance.
The instability issue is mainly related to the oxidation of Sn$^{2+}$ to Sn$^{4+}$ \cite{leijtens2017Sn_oxidation}.
The degradation usually starts from the surfaces or grain boundaries and is also sensitive to environmental factors including humidity and oxygen \cite{lanzetta2021degradation_CsSnI3}.
Insights into the surface properties will contribute to studying the degradation and defect formation mechanism to further improve the stability of the perovskites.
Experimentally various passivation methods have been proposed to improve the stability and quality of CsSnI$_3$ \cite{li2021cssni3_passivation}.
A recent scanning tunneling microscopy (STM) study of the CsSnI$_3$ (001) surface has reported surface reconstructions and Cs vacancy formation \cite{she2023cssni3_stm_001surface}.
Theoretically, structures and energies have been calculated for the (001) surface of orthorhombic CsSnI$_3$ \cite{jung2017ortho_cssni3_001}, the (100), (110), and (111) surfaces of cubic CsSnCl$_3$ \cite{chu2021cubic_cssncl3_surface}, and the (100), (110), and (111) surfaces of the cubic CsSnI$_3$ \cite{chen2023cubic_cssnx3_surface}.
However, theoretical investigation on the surface phase diagram, which is important to study surface stability at various chemical potentials for the orthorhombic phase of CsSnI$_3$ is still lacking. 
Additionally, DFT studies of the surface structures of Pb-based and Sn-based perovskites have been mainly focused on the conventional cells, ignoring the possibility of surface reconstructions.

In this work, we generate supercells of the (001), (110), and (100) surfaces of orthorhombic CsSnI$_3$ and determine the most stable surfaces under various Cs, Sn, and I chemical potentials.
We focus on these three surfaces as they are predicted to be more stable than other low-index surfaces in orthorhombic Pb-based perovskites by DFT calculations \cite{wang2015stoichiometric_MAPbI3, haruyama2014mapbi3_surface}.
The (001) surface has been observed experimentally by STM in MAPbI$_3$ \cite{she2016stmMAPbI3} and CsSnI$_3$ \cite{she2023cssni3_stm_001surface}.
X-ray diffraction measurements also reveal (110) and (001) peaks in both Sn-based and Pb-based perovskites \cite{liu2013mapbi3_xrd, li2024cssni3_xrd, dang2016xrd_fasni3}.
We create over 50 surface structures of CsSnI$_3$, ranging from SnI-rich, CsI-rich, I-rich, Sn-rich, Cs-rich, and stoichiometric surfaces and performed DFT calculations to investigate the atomic structures, surface energies, and electronic properties of these surfaces. 
In order to determine the most stable surfaces under different Cs, Sn, and I conditions, we calculate the surface energy using the grand potential approach and create surface phase diagrams to illustrate the most stable surfaces under different chemical potentials.
We find stable stoichiometric and CsI-terminated surfaces on (001), (110), and (100) under I-poor condition and several I-rich surfaces under high I chemical potentials. 
We discuss the electronic properties of these surfaces and propose growth conditions that results in clean surfaces with no mid-gap states and benefits device performance.

\section{\label{sec:level1}Computational Methods}

DFT calculations are performed using the Vienna Ab initio Simulation Package (VASP) \cite{kresse1996vasp1, kresse1996vasp2}.
We used the Perdew-Burke-Ernzerhof (PBE) functional \cite{Perdew_pbe1996} and a plane-wave energy cutoff of 400 eV.
The Brillouin zone of the orthorhombic unit cell of CsSnI$_3$ is sampled using a 2$\times$2$\times$2 {\bf k}-point grid.
The computed lattice constants of CsSnI$_3$ are $a$ = 8.66 $\mathrm{\AA}$, $b$ = 8.98 $\mathrm{\AA}$, $c$ = 12.52 $\mathrm{\AA}$, in reasonable agreement with the experimental values $a$ = 8.69 $\mathrm{\AA}$, $b$ = 8.64 $\mathrm{\AA}$, $c$ = 12.38 $\mathrm{\AA}$ \cite{cssni3_lattice_exp}.

\begin{figure}
\includegraphics[width=85mm]{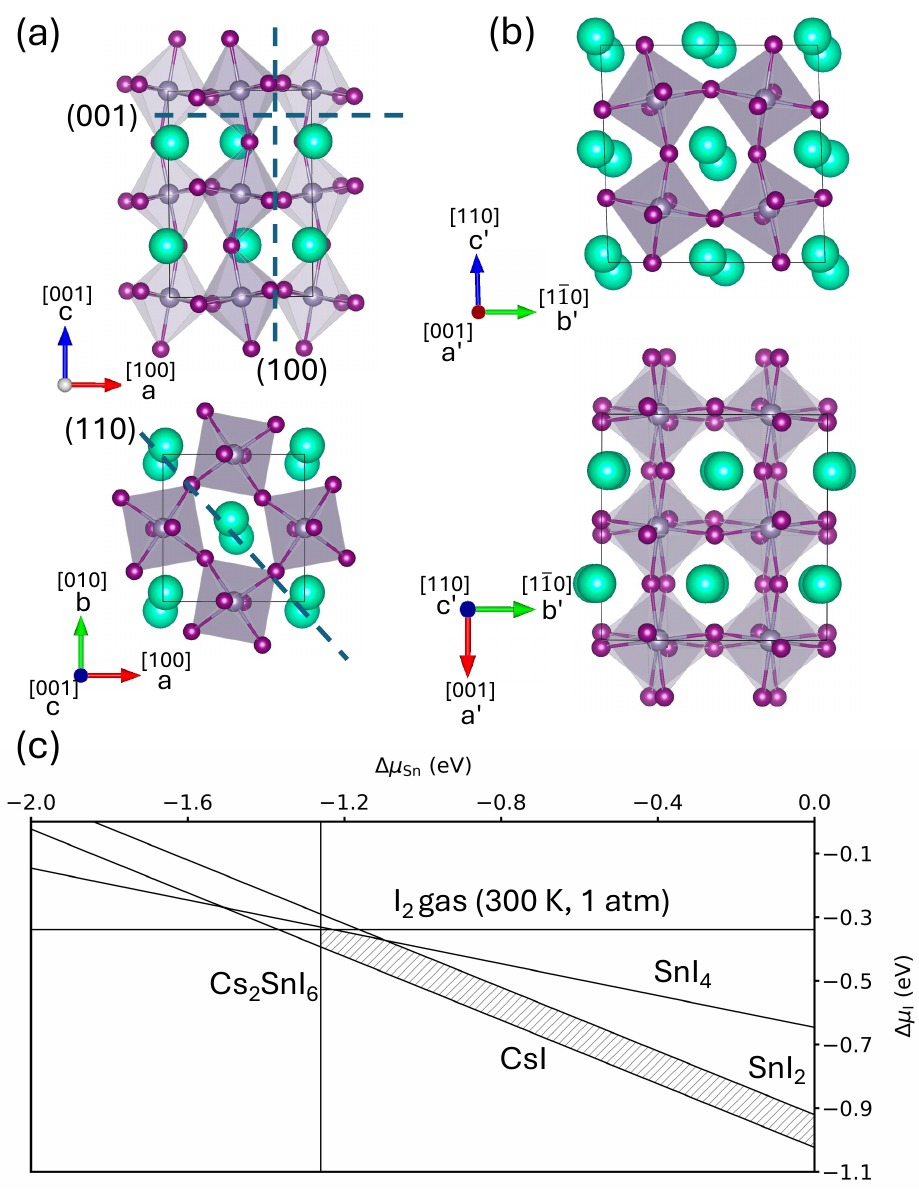}
\caption{\label{fig:chemical_potential} Top and side view of (a) the unit cell of CsSnI$_3$ and (b) the rotated unit cell of CsSnI$_3$. (c) The thermodynamically stable region for CsSnI$_3$, hatched in grey.
Color code: Cs (green), Sn (grey), and I (purple).
}
\end{figure}

Figure ~\ref{fig:chemical_potential} (a) shows the structure of the unit cell of orthorhombic CsSnI$_3$. Each unit cell contains four atomic layers along the [001] direction (two SnI$_2$ layers and two CsI layers) and two layers along the [100] direction (both mixed Cs-Sn-I layers).
The (001) slab is created by a 2$\times$2$\times$2.25 supercell, containing 9 layers along the [001] direction.
The (100) slab is created by a 3$\times$2$\times$1 supercell, containing 6 layers along the [100] direction.
To create the slab models for the (110) surface, a new unit cell is needed [Fig. ~\ref{fig:chemical_potential} (b)]: $a'=c$, $b'=a-b$, and $c'=a+b$.
The (110) surface slab models are created by a 1$\times$2$\times$2.25 supercell using the new unit cell, containing 9 layers along the [110] direction. 
These slabs are at least 25 {\AA} thick. The vacuum thickness is at least 20 {\AA}. We have performed convergence test of the slab and vacuum thickness: increasing the slab or vacuum thickness only changes the surface energy by less than 0.01 J/m$^2$. 
The inversion symmetry of the slabs allows identical surface structures on the top and bottom surfaces.
We use the equivalent {\bf k}-point grid for the supercell calculations as those adopted for the unit cell.
For example, the Brillouin zone of the (100) slab is sampled by a 1$\times$1$\times$2 {\bf k}-point grid.
Convergence tests on k-point sampling were also adopted for the bulk and surface systems. 
Changing the k-mesh to 4×4×3 for bulk and equivalently for slabs only changes the surface energy by less than 0.02 J/m$^2$ and does not change the surface phase diagrams.
All atoms in the slab are allowed to relax during the structural optimization until forces are smaller than 0.01 eV/{\AA}.

To compare the stability of surfaces with different stoichiometries, we created over 50 surface terminations of CsSnI$_3$, ranging from SnI$_2$-rich, CsI-rich, I-rich, Sn-rich, Cs-rich, and stoichiometric surfaces. 
We label the surface structures using the total number of atoms in the slab: Cs$_\alpha$Sn$_\beta$I$_\gamma$(CsSnI$_3$)$_\theta$, and use the grand potential approach \cite{haruyama2014mapbi3_surface,sanna2010grand_potential,qian1988gaas_grandpotential} to determine the surface stabilities.
The grand potential ($\Omega$) of the surface is defined as
\begin{multline} 
\Omega=\frac{1}{2A}\{E{_{\mathrm{tot}}}[\mathrm{Cs}_\alpha\mathrm{Sn}_\beta\mathrm{I}_\gamma\mathrm{(CsSnI}_3\mathrm{)}_\theta]-\alpha\mu_\mathrm{Cs}-\beta\mu_\mathrm{Sn}\\-\gamma\mu_\mathrm{I}-\theta\mu_\mathrm{CsSnI_3}\}.
\label{eq:grand_potential}
\end{multline}
$E_{\mathrm{tot}}$[Cs$_\alpha$Sn$_\beta$I$_\gamma$(CsSnI$_3$)$_\theta$] is the total energy of the slab.
$A$ is the surface area.
$\mu_\mathrm{Cs}$, $\mu_\mathrm{Sn}$, and $\mu_\mathrm{I}$ are defined as $\mu_\mathrm{Cs}=\mu_\mathrm{Cs}^\mathrm{bulk}+\Delta\mu_\mathrm{Cs}$, $\mu_\mathrm{Sn}=\mu_\mathrm{Sn}^\mathrm{bulk}+\Delta\mu_\mathrm{Sn}$, and $\mu_\mathrm{I}=\frac{1}{2}\mu_\mathrm{I_2}^\mathrm{gas}+\Delta\mu_\mathrm{I}$.
$\mu_\mathrm{CsSnI_3}$ is defined as $\mu_\mathrm{CsSnI_3}=\mu_\mathrm{CsSnI_3}^\mathrm{bulk}+\Delta\mu_\mathrm{CsSnI_3}$.
$\mu_\mathrm{Cs}^\mathrm{bulk}$ ($\mu_\mathrm{Sn}^\mathrm{bulk}$) is the single atom energy of the bulk Cs (Sn).
$\mu_\mathrm{I_2}^\mathrm{gas}$ is the energy of the I$_2$ molecule.
$\mu_\mathrm{CsSnI_3}^\mathrm{bulk}$ is the energy of the bulk CsSnI$_3$.
$\Delta\mu_\mathrm{CsSnI_3}$ is 0 under thermodynamic equilibrium condition.
$\Delta\mu_\mathrm{Cs}$, $\Delta\mu_\mathrm{Sn}$, and $\Delta\mu_\mathrm{I}$ are the chemical potentials of Cs, Sn, and I, constrained by the thermodynamic equilibrium condition of CsSnI$_3$ 
\begin{equation}
\Delta\mu_\mathrm{Cs}+\Delta\mu_\mathrm{Sn}+3\Delta\mu_\mathrm{I}=\Delta{H}_\mathrm{CsSnI_3},
\label{eq:delta_H_CsSnI3}
\end{equation}
and against the formation of the competing secondary phases of SnI$_2$, SnI$_4$, CsI, and Cs$_2$SnI$_6$: 
\begin{equation}
\begin{gathered} 
\Delta\mu_\mathrm{Sn}+2\Delta\mu_\mathrm{I}<\Delta{H}_\mathrm{SnI_2} (-1.84\: \mathrm{ eV}), \\ \Delta\mu_\mathrm{Sn}+4\Delta\mu_\mathrm{I}<\Delta{H}_\mathrm{SnI_4} (-2.59\: \mathrm{ eV}), \\
\Delta\mu_\mathrm{Cs}+\Delta\mu_\mathrm{I}<\Delta{H}_\mathrm{CsI} (-3.31\: \mathrm{ eV}), \\
2\Delta\mu_\mathrm{Cs}+\Delta\mu_\mathrm{Sn}+6\Delta\mu_\mathrm{I}<\Delta{H}_\mathrm{Cs_2SnI_6} (-9.46\: \mathrm{ eV}).
\label{eq:secondary_phases}
\end{gathered}
\end{equation}
The numbers in parentheses are the calculated formation enthalpy of the secondary phases.
Using the ideal gas model \cite{paudel2012ideal_gas,kang2017def_cspbbr3}, $\Delta\mu_\mathrm{I}$ is $-$0.34 eV at 300 K and 1 atm. 
The computational detail of $\Delta\mu_\mathrm{I}$ under ambient conditions is included in the Supplemental Material.
To avoid the formation of I$_2$, $\Delta\mu_\mathrm{I}$ is constrained to be $\Delta\mu_\mathrm{I}<-0.34$ eV.
The thermodynamically stable region is shown in Fig. ~\ref{fig:chemical_potential}(c).
The upper right and lower left borders represent SnI$_2$ and CsI-rich conditions.

The formation enthalpy of CsSnI$_3$ ($\Delta{H}_\mathrm{CsSnI_3}$) is calculated to be $-5.36$ eV.
Substituting the $\Delta\mu_\mathrm{Cs}$ with $\Delta\mu_\mathrm{Cs}=\Delta{H}_\mathrm{CsSnI_3}-\Delta\mu_\mathrm{Sn}-3\Delta\mu_\mathrm{I}$ in Equation~\ref{eq:grand_potential}, the grand potential can be defined as a function of $\Delta\mu_\mathrm{Sn}$ and $\Delta\mu_\mathrm{I}$:
\begin{multline} 
\Omega=\frac{1}{2A}\{E{_{\mathrm{tot}}}[\mathrm{Cs}_\alpha\mathrm{Sn}_\beta\mathrm{I}_\gamma\mathrm{(CsSnI}_3\mathrm{)}_\theta]-\alpha(\Delta{H}_\mathrm{CsSnI_3}\\-\Delta\mu_\mathrm{Sn}-3\Delta\mu_\mathrm{I}+\mu_\mathrm{Cs}^\mathrm{bulk})-\beta(\Delta\mu_\mathrm{Sn}+\mu_\mathrm{Sn}^\mathrm{bulk})-\gamma(\Delta\mu_\mathrm{I}+\frac{1}{2}\mu_\mathrm{I_2}^\mathrm{gas})\\-\theta\mu_\mathrm{CsSnI_3}^\mathrm{bulk}\}.
\label{eq:formation_energy2}
\end{multline}
We then calculated $\Omega$ for all the relaxed surface structures, and used our in-house code to generate the surface phase diagrams.   
In the surface phase diagrams, each shaded region represents the surface structure with the lowest $\Omega$ under a certain combination of Sn and I chemical potentials.
We will mainly focus on discussing the stable surface structures near the thermodynamically stable region of CsSnI$_3$ [hatched in grey in Fig.~\ref{fig:chemical_potential}(c)].

\section{\label{results}Results and discussions}
We will discuss the phase diagram and stable surfaces on (001) (section \ref{sec001}), (110) (section \ref{sec110}), and (100) (section \ref{sec100}) under the thermodynamically stable regions of CsSnI$_3$. In section \ref{combined}, we generate a surface phase diagrams of all three terminations and discuss the electronic properties and growth conditions.

\begin{figure}
\includegraphics[width=84mm]{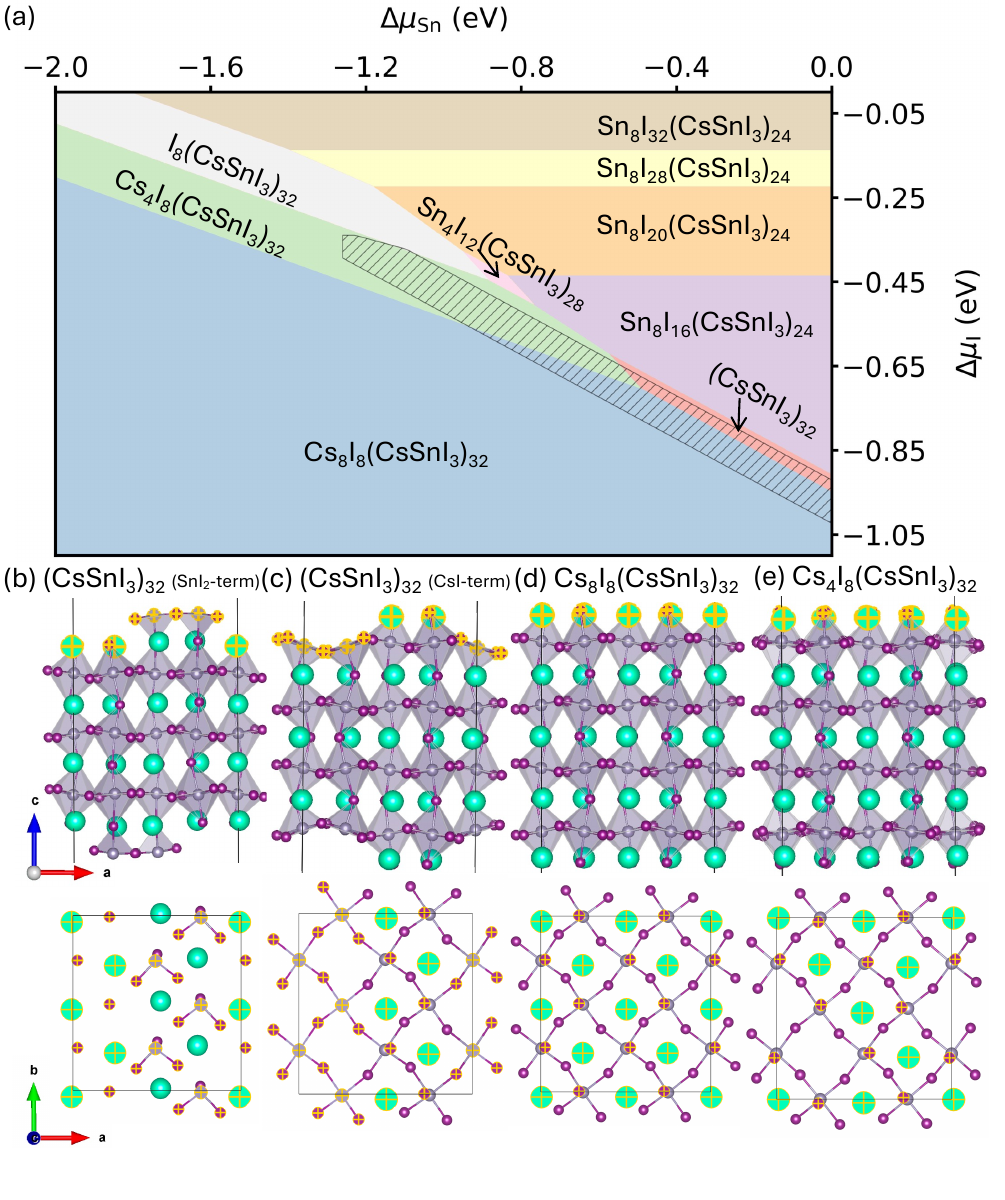}
\caption{\label{fig:001} (a) Surface phase diagram of the CsSnI$_3$ (001) surface under different Sn and I chemical potentials.
The atomic structures of the surfaces that overlap with the thermodynamically stable region of CsSnI$_3$ (hatched in grey): (b) (CsSnI$_3$)$_{32}$ (SnI$_2$-terminated), (c) (CsSnI$_3$)$_{32}$ (CsI-terminated), (d) Cs$_{8}$I$_{8}$(CsSnI$_3$)$_{32}$, and (e) Cs$_{4}$I$_{8}$(CsSnI$_3$)$_{32}$.
Color code is the same as Fig. ~\ref{fig:chemical_potential} and surface atoms on the top layer are highlighted.
}
\end{figure}

\subsection{\label{sec001} (001) surface}

We created surfaces with 20 different stoichiometries for (001).
For certain stoichiometries, we generated several structures to obtain the most stable configuration and in total 30 structures were calculated.
The grand potentials of the 20 surfaces with different stoichiometries are plotted as a function of $\Delta\mu_\mathrm{Sn}$ in Fig. S1 in the Supplemental Material.
On the (001) surface, we find three stable surfaces in the thermodynamically stable region of CsSnI$_3$.
Under I-poor conditions ($\Delta\mu_\mathrm{I}<-$0.65 eV), the surface prefers the stoichiometric surface with (CsSnI$_3$)$_{32}$ [Fig.~\ref{fig:001}(b)] near the SnI$_2$ border, and prefers the Cs$_{8}$I$_{8}$(CsSnI$_3$)$_{32}$ surface [Fig.~\ref{fig:001}(d)] near the CsI border.
Moving to a more SnI$_2$-rich region, the SnI$_2$-terminated flat surface [Sn$_{8}$I$_{16}$(CsSnI$_3$)$_{24}$] is stable in the phase diagram. We note that this surface is outside the thermodynamically stable region of CsSnI$_3$.
The flat CsI-terminated CsSnI$_3$ (001) surface has been observed experimentally by scanning tunneling microscopy \cite{she2023cssni3_stm_001surface}, which is consistent with our observation of the flat CsI-terminated surface [Cs$_{8}$I$_{8}$(CsSnI$_3$)$_{32}$] near the CsI-border in the phase diagram.
We note that the DFT study on the CsPbI$_3$ (001) surface also shows that the CsI-terminated surface is more stable than the PbI$_2$-terminated surface \cite{seidu2021cspbi3001_dft}.

The (CsSnI$_3$)$_{32}$ surface shown in Fig.~\ref{fig:001}(b) is a stoichiometric surface with $\Omega$ calculated to be 1.36 eV/cell (0.07 J/m$^2$).
This stoichiometric surface is generated by removing four Sn and eight I atoms on both top and bottom surfaces of the Sn$_{8}$I$_{16}$(CsSnI$_3$)$_{24}$ surface and is therefore terminated with the SnI$_2$ layer.
Interestingly, we also found another stoichiometric surface [Fig.~\ref{fig:001}(c)] with a slightly higher energy ($\Omega=$ 0.08 J/m$^2$).
This surface is created by removing four Cs and four I atoms from both top and bottom surfaces from the Cs$_{8}$I$_{8}$(CsSnI$_3$)$_{32}$ surface, and is terminated with the CsI layer.
As these two surfaces are close in energy, we predict that they will co-exist on CsSnI$_3$(001).

Moving to a higher I chemical potential, we find several I-rich surfaces. 
Under Sn-poor region, the surface with Cs$_{4}$I$_{8}$(CsSnI$_3$)$_{32}$ [Fig.~\ref{fig:001}(e)] is stable in the thermodynamically stable region of CsSnI$_3$.
Under the SnI$_2$-rich conditions, we find the surfaces with Sn$_{8}$I$_{20}$(CsSnI$_3$)$_{24}$, Sn$_{8}$I$_{28}$(CsSnI$_3$)$_{24}$, and Sn$_{8}$I$_{32}$(CsSnI$_3$)$_{24}$ to be stable: the number of I atoms will increase as the chemical potential of I increases.
These SnI$_{2}$ or I-rich surfaces do not fall into the thermodynamically stable region of CsSnI$_3$.

We note that it is important to use a supercell to study the surface structures of CsSnI$_3$, as reconstructions may occur.
On the (CsSnI$_3$)$_{32}$ surfaces [Figs~\ref{fig:001}(b-c)], we find the most stable surface under this stoichiometry is the 2$\times$1 reconstructed surface. 
To study the I-rich surfaces, the 2$\times$2 supercell allows a more diverse combination of the Cs, Sn, and I ratios, and indeed, we find the Cs$_{4}$I$_{8}$(CsSnI$_3$)$_{32}$ surface to be stable in the surface phase diagram, which is a 2$\times$2 reconstructed surface.

\subsection{\label{sec110} (110) surface}

\begin{figure}
\includegraphics[width=84mm]{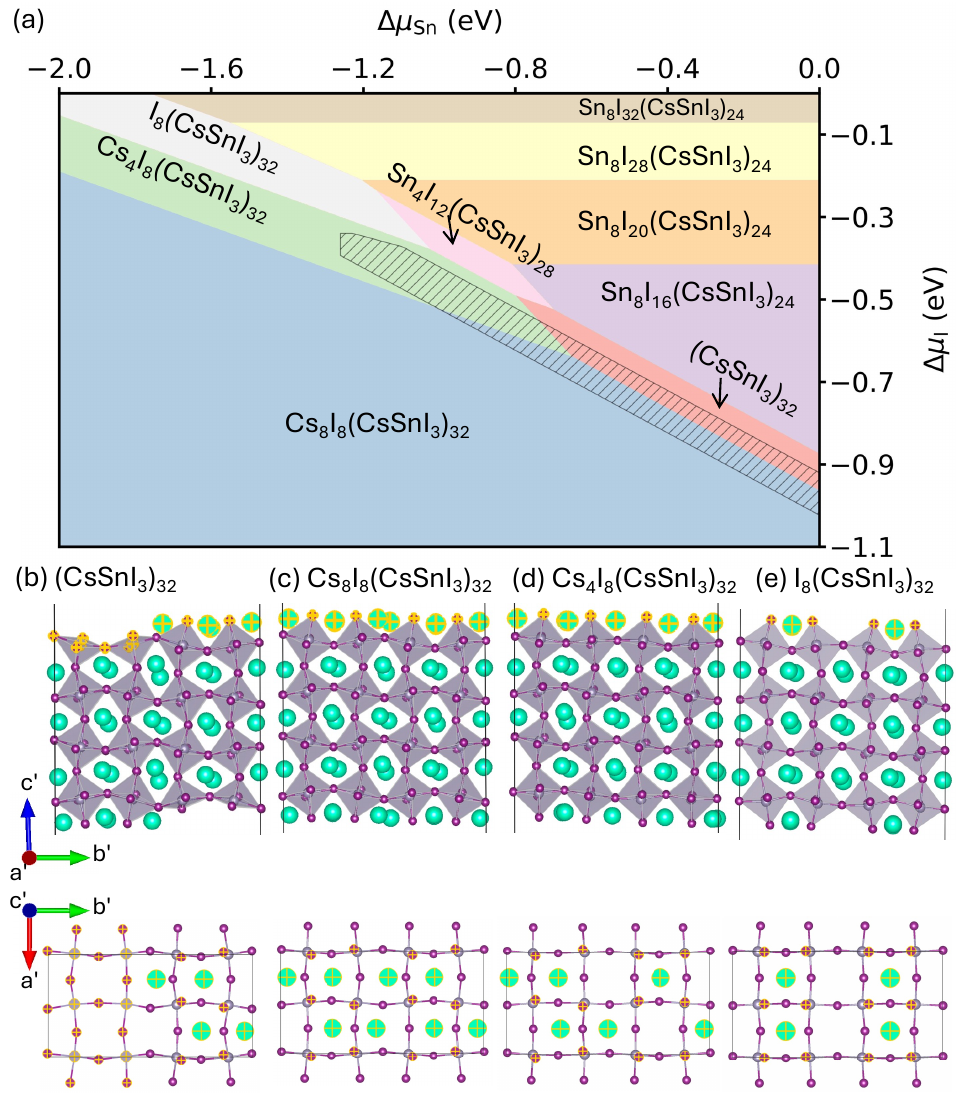}
\caption{\label{fig:110} (a) Surface phase diagram of the CsSnI$_3$ (110) surface under different Sn and I chemical potentials.
The atomic structures of (b) (CsSnI$_3$)$_{32}$, (c) Cs$_{8}$I$_{8}$(CsSnI$_3$)$_{32}$, (d) Cs$_{4}$I$_{8}$(CsSnI$_3$)$_{32}$, and (e) I$_{8}$(CsSnI$_3$)$_{32}$.
}
\end{figure}

Similar to the (001) surface, the (110) surface slab also consists of the alternating SnI$_2$ and CsI layers. These two surfaces have very similar bonding environment.
We created surfaces with 10 different stoichiometries for (110), and generated 15 structures, considering reconstructions for certain stoichiometries.
The grand potentials of the 10 surfaces with different stoichiometries are also plotted as a function of $\Delta\mu_\mathrm{Sn}$ in Fig. S1 in the Supplemental Material.
The surface phase diagram of (110) [Fig.~\ref{fig:110}(a)] is close to that of (001) [Fig.~\ref{fig:001}(a)].
It is also important to use a supercell to calculate the (110) surfaces, as we find most of the surface terminations have the 2$\times$1 reconstruction.
The supercell is generated by expanding the cell size along the [1$\bar{1}$0] ($b'$) direction using the rotated unit cell shown in Fig.~\ref{fig:chemical_potential}(b).

Under the I-poor conditions, we find that the stoichiometric (CsSnI$_3$)$_{32}$ surface [Fig.~\ref{fig:110}(b)] is stable near the SnI$_2$ border and the Cs$_{8}$I$_{8}$(CsSnI$_3$)$_{32}$ flat surface [Fig.~\ref{fig:110}(d)] is stable near the CsI border.
The (CsSnI$_3$)$_{32}$ slab is generated by removing eight Cs and eight I atoms from the slab with Cs$_{8}$I$_{8}$(CsSnI$_3$)$_{32}$ (removing four Cs and four I atoms from both the top and bottom flat surface).
Compared to the (CsSnI$_3$)$_{32}$ surface on (001), the (CsSnI$_3$)$_{32}$ surface on (110) is slightly lower in energy: 1.12 eV/cell (0.06 J/m$^2$).

We explain the lower energy on (110) by comparing the number of Sn-I bonds and Cs-I bonds that need to be broken to form the (CsSnI$_3$)$_{32}$ surfaces on (001) and (100), which is summarized in Table ~\ref{tab:grand_potential}.
In bulk CsSnI$_3$, Cs is bonded in an 8-coordinate geometry, and there is a spread of Cs–I bond distances ranging from 3.95–4.22 \AA. Sn is bonded to six I atoms to form corner-sharing SnI$_6$ octahedra with the Sn-I distances ranging from 3.18 to 3.20 \AA.
The number of broken bonds on the surfaces is then counted assuming that Cs is 8-fold coordinated and Sn is 6-fold coordinated.
To form the (CsSnI$_3$)$_{32}$ surface, there are 8 Sn-I bonds and 24 Cs-I bonds broken on (001), and 8 Sn-I bonds and 20 Cs-I bonds broken on (110).
Additionally, the (110) surface has a slightly larger surface area, explaining the slightly lower surface energy of (110).

Under I-rich conditions, the surface with Cs$_{4}$I$_{8}$(CsSnI$_3$)$_{32}$ [Fig.~\ref{fig:110}(d)] is stable in the thermodynamically stable region of CsSnI$_3$.
The surfaces with I$_{8}$(CsSnI$_3$)$_{32}$ [Fig.~\ref{fig:110}(e)] and Sn$_{4}$I$_{12}$(CsSnI$_3$)$_{28}$ are stable under a more Sn-rich condition.
This is also consistent with the (001) surface due to the similar surface structure and bonding environment.

\begin{figure}
\includegraphics[width=84mm]{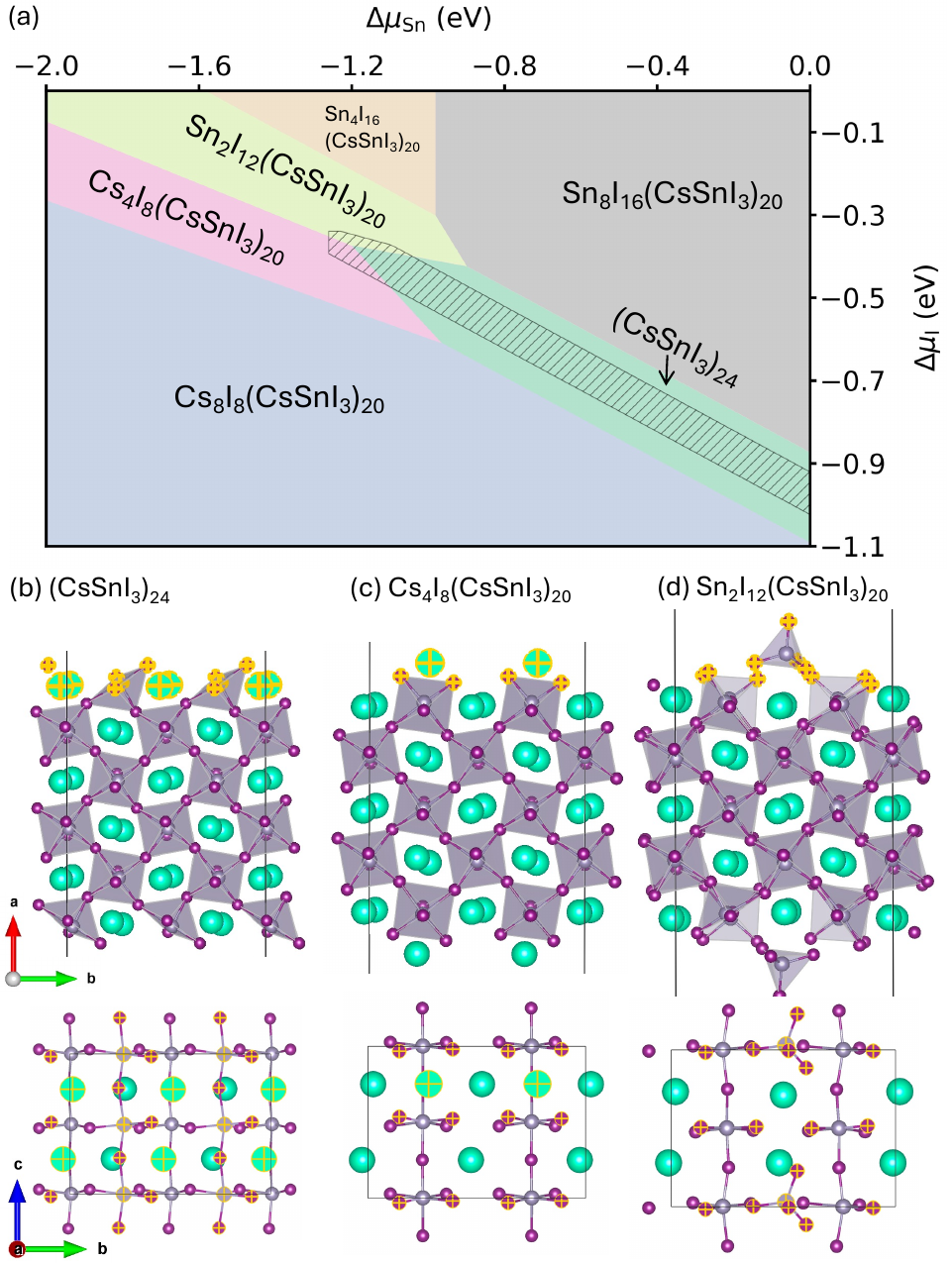}
\caption{\label{fig:100} (a) Surface phase diagram of the CsSnI$_3$ (100) surface under different Sn and I chemical potentials.
The atomic structures of the surfaces that overlap with the thermodynamically stable region of CsSnI$_3$ (hatched in grey): (b) (CsSnI$_3$)$_{24}$, (c) Cs$_{4}$I$_{8}$(CsSnI$_3$)$_{20}$, and (d) Sn$_{2}$I$_{12}$(CsSnI$_3$)$_{20}$.
}
\end{figure}

\subsection{\label{sec100} (100) surface}

The (100) surface has a very different bonding environment compared to (001) and (110): each layer is a combined Cs-Sn-I layer. 
By cutting different number of Sn-I and Cs-I bonds, we can also generate stoichiometric surfaces, CsI-rich, SnI$_2$-rich, and I-rich surfaces.
We created surfaces with 10 different stoichiometries, and generated 12 structures, considering reconstructions for two stoichiometries.
The grand potentials of the 10 surfaces with different stoichiometries are plotted as a function of $\Delta\mu_\mathrm{Sn}$ in Fig. S1 in the Supplemental Material.
Figure ~\ref{fig:100}(a) shows that the thermodynamically stable region of CsSnI$_3$ is mostly located on the stoichiometric surface (CsSnI$_3$)$_{24}$ [Fig.~\ref{fig:100}(b)], indicating the predominance of this surface. 
This surface has the lowest energy (0.62 eV/cell = 0.04 J/m$^2$) compared to (001) and (110).
This is an unreconstructed surface, and it is stable under a very wide range of Sn and I chemical potentials.

We also summarize the number of bonds that need to be broken to create the (100) (CsSnI$_3$)$_{24}$ surface in Table ~\ref{tab:grand_potential} to compare the energies of different surface terminations.
To create the stoichiometric (CsSnI$_3$)$_{32}$ surfaces on (110), we need to break 8 Sn-I bonds and 20 Cs-I bonds in each supercell, corresponding to 0.026 Sn-I bonds/$\mathrm{\AA}^2$ and 0.064 Cs-I bonds/$\mathrm{\AA}^2$.
To generate the (CsSnI$_3$)$_{24}$ surface on (100), there are more Sn-I (0.036/$\mathrm{\AA}^2$) and fewer Cs-I (0.053/$\mathrm{\AA}^2$) bonds to break.
We also note that the energies of the unrelaxed (110) (CsSnI$_3$)$_{32}$ and (100) (CsSnI$_3$)$_{24}$ surfaces are similar.
The lower energy on the (100) (CsSnI$_3$)$_{24}$ surface can thus be explained by the relaxation effect.

Moving to more CsI-rich conditions, we also find the CsI-rich surface [Cs$_8$I$_8$(CsSnI$_3$)$_{20}$] in the phase diagram. 
Different from the (001) and (001) terminations, this CsI-rich surface is not stable in the thermodynamically stable region of CsSnI$_3$.
Under more I-rich conditions, there are two I-rich surfaces that overlap with the thermodynamically stable region of CsSnI$_3$: Cs$_{4}$I$_{8}$(CsSnI$_3$)$_{20}$ [Fig.~\ref{fig:100}(c)] and Sn$_{2}$I$_{12}$(CsSnI$_3$)$_{20}$ [Fig.~\ref{fig:100}(d)]. 
However, we note that they are only stable in a very narrow region of the thermodynamically stable region of CsSnI$_3$.

\subsection{\label{combined} Surface phase diagram of (001), (110), and (100) and electronic properties}

\begin{figure}
\includegraphics[width=85mm]{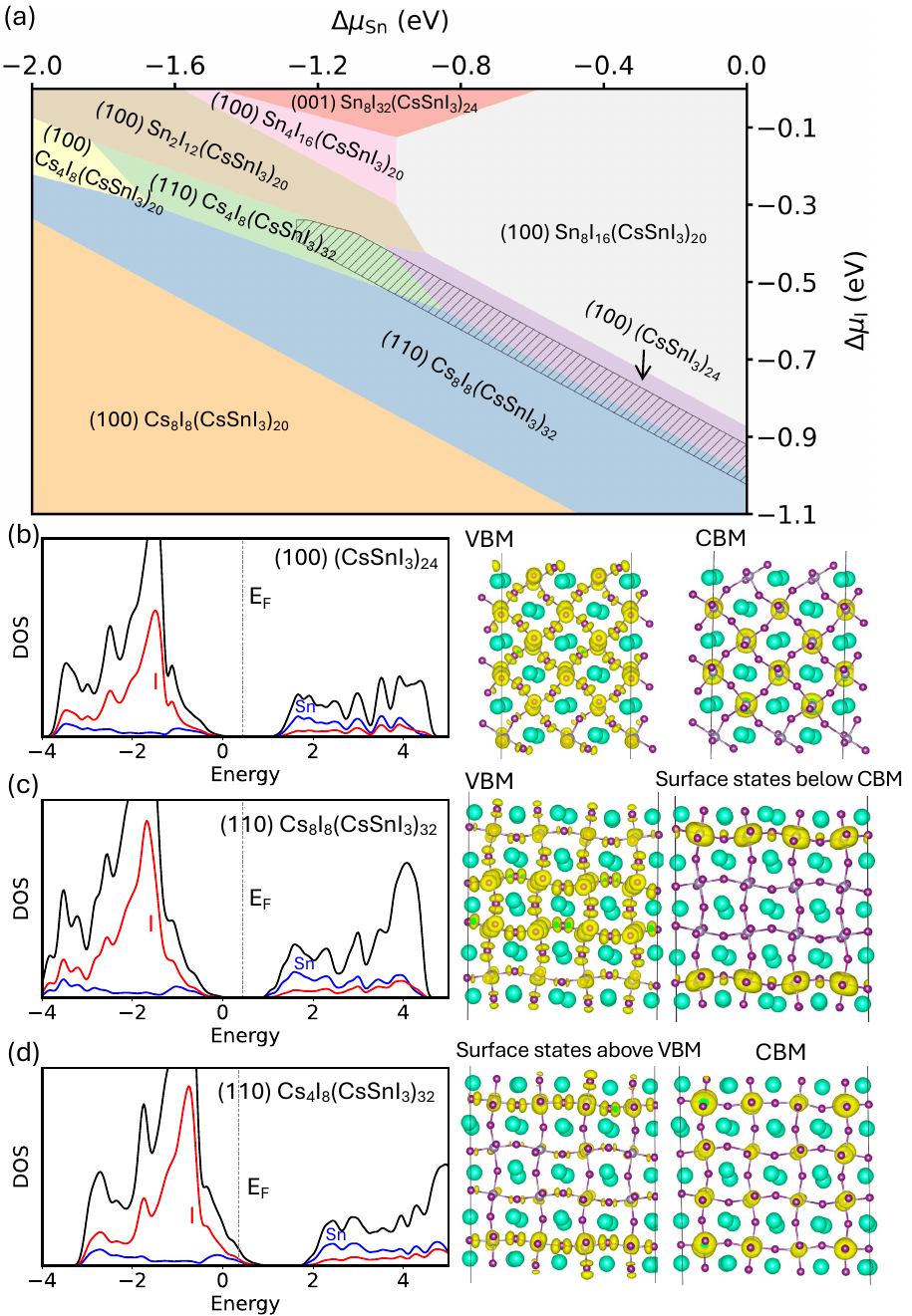}
\caption{\label{fig:phase_diagram} (a) Surface phase diagram of the CsSnI$_3$ (001), (110), and (100) under different Sn and I chemical potentials. The density of states and partial charge densities of VBM, CBM, or surface states of the (b) (100)-(CsSnI$_3$)$_{24}$, (c) (110)-Cs$_{8}$I$_{8}$(CsSnI$_3$)$_{32}$, and (d) (110)-Cs$_{4}$I$_{8}$(CsSnI$_3$)$_{32}$ surfaces.
}
\end{figure}

To compare the stabilities of all three surfaces under different chemical potentials, we generate the surface phase diagram including all stable surfaces in Fig.~\ref{fig:001}(a), Fig.~\ref{fig:110}(a), and Fig.~\ref{fig:100}(a).
The phase diagram shown in Fig.~\ref{fig:phase_diagram} (a) indicates that (100) surface is the most stable surface under a wide range of Sn and I chemical potentials in the thermodynamically stable region of CsSnI$_3$.
The (110)-Cs$_{8}$I$_{8}$(CsSnI$_3$)$_{32}$ surface is stable in narrower region close to the CsI-border. 
Moving to more I-rich conditions, the (110)-Cs$_{4}$I$_{8}$(CsSnI$_3$)$_{32}$ surface becomes more stable. 

\begin{table}[htb]
\caption{\label{tab:grand_potential}%
The grand potential ($\Omega$ in eV/cell), surface area ($A$ in $\mathrm{\AA}^2$), and the number of bonds broken on different surfaces in /cell (/$\mathrm{\AA}^2$).}
\begin{ruledtabular}
\begin{tabular}{>{\centering}p{0.27\linewidth}|p{0.12\linewidth}|p{0.12\linewidth}|p{0.14\linewidth}|p{0.16\linewidth}}
\textrm{ }&
\textrm{$\Omega$}&
\textrm{$A$}&
\textrm{Sn-I} &
\textrm{Cs-I} \\
\colrule
(001)-(CsSnI$_3$)$_{32}$ & 1.48 & 311.07 &  8 (0.026) & 24 (0.077) \\
\hline 
(001)-Cs$_{8}$I$_{8}$(CsSnI$_3$)$_{32}$ & -12.48-4$\Delta\mu_\mathrm{Cs}$-4$\Delta\mu_\mathrm{I}$ & 311.07  & 8 (0.026) & 24 (0.077) \\
\hline 
(110)-(CsSnI$_3$)$_{32}$ & 1.12 & 312.37 &  8 (0.026) & 20 (0.064) \\
\hline 
(110)-Cs$_{8}$I$_{8}$(CsSnI$_3$)$_{32}$ & -12.60-4$\Delta\mu_\mathrm{Cs}$-4$\Delta\mu_\mathrm{I}$ & 312.37 & 8 (0.026) & 20 (0.064)  \\
\hline 
(100)-(CsSnI$_3$)$_{24}$ & 0.62 & 224.74 & 8 (0.036) & 12 (0.053)\\
\end{tabular}
\end{ruledtabular}
\end{table}

To study the effect of chemical potential and surface terminations on the electronic structure of the CsSnI$_3$ surfaces, we show the projected density of states (PDOS) of the (100)-(CsSnI$_3$)$_{24}$ [Fig.~\ref{fig:phase_diagram} (b)], (110)-Cs$_{8}$I$_{8}$(CsSnI$_3$)$_{32}$ [Fig.~\ref{fig:phase_diagram} (c)], and (110)-Cs$_{4}$I$_{8}$(CsSnI$_3$)$_{32}$ surfaces [Fig.~\ref{fig:phase_diagram} (d)].
The red and blue curves are the projected density of states of the I and Sn atoms.
Consistent with previous DFT studies, the valence band consists mainly of the I $p$ states while the conduction band is mainly contributed by Sn $p$ states \cite{dalpian2017cssni3dos}.

Figure ~\ref{fig:phase_diagram} (b) shows that the stoichiometric (100)-(CsSnI$_3$)$_{24}$ surface has no localized surface states in the bandgap.
We also plot the charge distribution of the states at the valence band maximum (VBM) and conduction band minimum (CBM), which show that both states are all delocalized bulk states.
On the (110)-Cs$_{8}$I$_{8}$(CsSnI$_3$)$_{32}$ surface, we find four surface states below the CBM, which are localized on the Sn atoms in the sub-surface. 
These surface states are 0.09 eV lower than the CBM, and will cause a slight shrink of the bandgap.

The spin-orbit coupling (SOC) effects on PDOS and the partial charge densities are also examined, as shown in Fig. S2 in the Supplemental Material.
The calculated bandgap of bulk CsSnI$_3$ is 0.88 eV using the PBE functional and 0.57 eV when SOC is included, which underestimates the experimental bandgap (1.3 eV) \cite{yu2011cssni3_bandgap}.
The smaller bandgap including SOC is due to the Sn-$p$ character of the conduction bands, which is consistent with previous DFT calculations \cite{su2021pbesoc_gap}.
We find a similar SOC effect on the bandgap of the slabs: the calculated bandgap using the PBE functional is 1.19 eV for the (100)-(CsSnI$_3$)$_{24}$ surface and 1.0 eV for the (110)-Cs$_{8}$I$_{8}$(CsSnI$_3$)$_{32}$ surface. 
The bandgaps are smaller when SOC is included: 0.97 eV for (100)-(CsSnI$_3$)$_{24}$ and 0.75 eV for (110)-Cs$_{8}$I$_{8}$(CsSnI$_3$)$_{32}$. 
We also note that due to the quantum confinement effect, the bandgap of a slab is slightly larger than bulk CsSnI$_3$ \cite{katan2019quantum_confinement}.
The partial charge densities of the VBM, CBM, and surface states using PBE+SOC (Fig. S2 in the Supplemental Material) are consistent with the PBE calculations.

Under I-rich conditions, the (110)-Cs$_{4}$I$_{8}$(CsSnI$_3$)$_{32}$ surface becomes stable, which has partially occupied states above VBM.
The partial charge density of the surface states above VBM shows that these surface states are localized on the surface and sub-surface I atoms.
This is consistent with previous DFT calculations on the surfaces with Cs vacancies that shows a shrink of the bandgap \cite{she2023cssni3_stm_001surface}.
The SOC effect on PDOS and the partial charge densities is also examined for the (110)-Cs$_{4}$I$_{8}$(CsSnI$_3$)$_{32}$ surface, as shown in Fig. S2 in the Supplemental Material.
The conduction band is slightly lower than the DOS calculated by PBE due to the SOC effect and the partial charge densities of the CBM and surface states using PBE+SOC are consistent with the PBE calculations.
We also note that these surface states may affect the hole capture process in nonradiative recombinations caused by the defects near or at the surface, and affect the charge carrier lifetimes.
From the phase diagram and the electronic structure analysis, we conclude that synthesizing the CsSnI$_3$ under I-poor (Sn-rich) conditions is beneficial to achieve a clean surface with fewer or no surface states, which contribute to long charge carrier lifetimes.
We note that SnCl$_2$ or SnF$_2$ has been used as additives to CsSnI$_3$-based photovoltaics, which can provide a more Sn-rich environment, and these additives have been experimentally proven to improve device stability and efficiency \cite{marshall2016sncl2additive}.

There are experimental evidence that the (001) surfaces of the orthorhombic MAPbI$_3$ are terminated by the MAI layer \cite{she2016stmMAPbI3}.
Recent STM study has also reported that the (001) surface of orthorhombic CsSnI$_3$ is terminated by the CsI layer \cite{she2023cssni3_stm_001surface}.
What is unique about the CsSnI$_3$ is that ordered Cs vacancies are observed on the (001) surface, leading to a double-chain or tetramer phase \cite{she2023cssni3_stm_001surface}.
These experimental observations are consistent with our phase diagram of the (001) surface where the Cs$_{8}$I$_{8}$(CsSnI$_3$)$_{32}$ and Cs$_{4}$I$_{8}$(CsSnI$_3$)$_{32}$ surfaces are stable under the thermodynamically stable region of CsSnI$_3$.
We note that the surfaces with Cs vacancies are charge neutral surfaces.

\begin{figure}
\includegraphics[width=85mm]{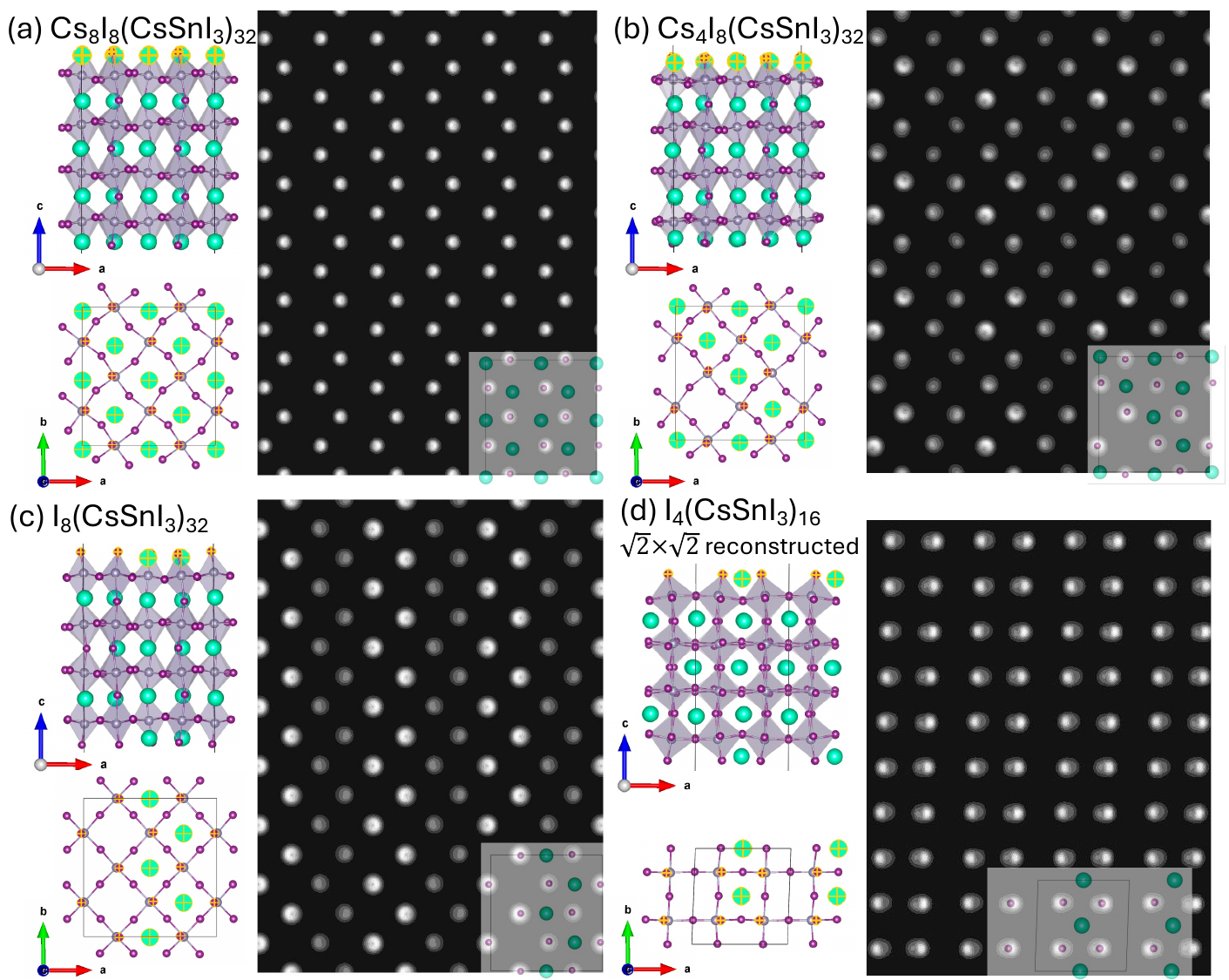}
\caption{\label{fig:stm} The simulated STM image of (a) Cs$_{8}$I$_{8}$(CsSnI$_3$)$_{32}$, (b) Cs$_{4}$I$_{8}$(CsSnI$_3$)$_{32}$, (c) I$_{8}$(CsSnI$_3$)$_{32}$, and (d) I$_{8}$(CsSnI$_3$)$_{32}$ ($\sqrt{2}\times\sqrt{2}$ reconstructed).
}
\end{figure}

To understand how Cs vacancies affect the STM images, we simulated several (001) surfaces with different Cs vacancy concentrations.
Figure ~\ref{fig:stm} shows the side and top view of the structures and the simulated STM images.
The STM images are obtained by integrating the density of states from $-$1.5 eV to 0 eV (Fermi level) using the constant current mode from post-processing VASPKIT package \cite{wang2021vaspkit}.
As the states near VBM are mainly contributed by the I $p$ states, the simulated STM image of the Cs$_{8}$I$_{8}$(CsSnI$_3$)$_{32}$ [Fig.~\ref{fig:stm}(a)] surface shows bright iodine atoms forming a zigzag pattern while the cesium atoms are invisible, which agrees well with the previous experimental and simulated STM results \cite{she2023cssni3_stm_001surface}.

Fig.~\ref{fig:stm}(b) and (c) show how Cs vacancies affect the STM images.
The Cs$_{4}$I$_{8}$(CsSnI$_3$)$_{32}$ surface [Fig.~\ref{fig:stm}(b)] is created by removing four Cs atoms from the flat surface Cs$_{8}$I$_{8}$(CsSnI$_3$)$_{32}$ (two Cs atoms are removed from the top layer and two Cs atoms are removed from the bottom layer).
Protrusions of the I atoms near the Cs vacancies are observed while the I atoms that are bonded to Cs atoms are darker in signal.
When four Cs atoms are removed from the surface with Cs$_{4}$I$_{8}$(CsSnI$_3$)$_{32}$, the surface termination becomes I$_{8}$(CsSnI$_3$)$_{32}$. 
The STM image of the I$_{8}$(CsSnI$_3$)$_{32}$ [Fig.~\ref{fig:stm}(c)] shows that the distance between the two I columns that are close to the Cs vacancies is longer than the two columns with Cs-I bonds.
In contrast to the zigzag pattern from the CsI-terminated surface, the surfaces with Cs vacancies lead to the tetrameric phase due to the change of the I-I distances.
We also simulated the STM image of the I$_{4}$(CsSnI$_3$)$_{16}$ surface, which is a $\sqrt{2}\times\sqrt{2}$ reconstructed surface with equivalent stoichiometry as I$_{8}$(CsSnI$_3$)$_{32}$.
The two columns of the Cs vacancies result in a change of the I-I bond length change, resulting in the double-chain pattern shown in Fig.~\ref{fig:stm}(d), which is consistent with the experimentally observed pattern on the CsSnI$_3$ (001) surface \cite{she2023cssni3_stm_001surface}.
These two surface structures are close in energy: at $\Delta\mu_\mathrm{I}=0$, $\Omega$ is -0.04 J/m$^2$ for I$_{8}$(CsSnI$_3$)$_{32}$ and -0.03 J/m$^2$ for the $\sqrt{2}\times\sqrt{2}$ reconstructed I$_{4}$(CsSnI$_3$)$_{16}$ surface.
We predict that these two surfaces can co-exist on CsSnI$_3$ (001).

\section{\label{conclution}Conclusions}

In conclusion, we perform DFT calculations to investigate the structure and energetics of the (001), (110), and (100) surfaces of orthorhombic CsSnI$_3$.
The CsSnI$_3$ (001) and (110) surfaces have similar bonding environment and surface reconstructions: under I-poor conditions, the unreconstructed CsI-terminated surface and the 2$\times$1 reconstructed stoichiometric surface are stable; under I-rich conditions, the surface prefers a I-rich surface with Cs$_{4}$I$_{8}$(CsSnI$_3$)$_{32}$ termination.
On the CsSnI$_3$ (100) surface, we find the unreconstructed stoichiometric surface to be the most stable under a wide range of I chemical potentials, and two I-rich surfaces are found under higher I chemical potentials. 

From the surface phase diagram of all surfaces, we predict that the stoichiometric (100) surface will be most likely exposed from the bulk-growth methods under I-poor conditions due to the low formation energy, which is illustrated by its high coverage of the thermodynamically stable region of CsSnI$_3$.
I-poor conditions will result in Cs vacancy formation, leading to I-rich surfaces that will change the electronic properties of the surface.
The I-rich surfaces result in partially occupied surface states above VBM and will affect charge carrier transport properties.
The stoichiometric surfaces of (001), (110), and (100) surfaces are stable under I-poor conditions and do not have surface states in the band gap while the CsI-terminated (001) and (110) surfaces only have a few surface states below CBM.
We conclude that the I-poor (Sn-rich) conditions is beneficial to achieve a clean surface with fewer or no surface states, which contribute to improved charge carrier transport properties.

\begin{acknowledgments}
The work was supported by the new faculty start-up funding from State University of New York at Binghamton.
Computing resources were provided by Theory and Computation at Center for Functional Nanomaterials, which is a U.S. Department of Energy Office of Science User Facility, and the Scientific Data and Computing Center, a component of the Computational Science Initiative, at Brookhaven National Laboratory, which are supported by the U.S. Department of Energy, Office of Basic Energy Sciences, under Contract No. DE-SC0012704.
Computing resources were also provided by the Spiedie cluster at State University of New York at Binghamton.

\end{acknowledgments}

\nocite{*}

\bibliography{Surface_phase_diagram_CsSnI3}

\end{document}


\title{Supplemental Material  \protect\\  Surface phase diagram of CsSnI$_3$ from first-principles calculations}

\author{Kejia Li}
\affiliation{Department of Electrical and Computer Engineering, State University of New York at Binghamton, Binghamton, New York 13902, United States}
\affiliation{Materials Science and Engineering Program, State University of New York at Binghamton, Binghamton, New York 13902, United States}
\author{Chadawan Khamdang}
\affiliation{Department of Electrical and Computer Engineering, State University of New York at Binghamton, Binghamton, New York 13902, United States}
\author{Mengen Wang}
\email{mengenwang@binghamton.edu}
\affiliation{Department of Electrical and Computer Engineering, State University of New York at Binghamton, Binghamton, New York 13902, United States}
\affiliation{Materials Science and Engineering Program, State University of New York at Binghamton, Binghamton, New York 13902, United States}

\maketitle

\pagebreak

Computational details on I$_2$ chemical potential\\

Using an ideal gas model, $\Delta\mu_\mathrm{I}$ is written as a function of pressure (P) and temperature (T) as

\begin{equation}
\Delta\mu_\mathrm{I}=[H_0+c_p(T-T_0)-TS_0+Tc_pln(T/T_0)+k_BTln(P/P_0)]/2.
\end{equation}

$T_0=$ 298 K and $P_0=$ 0.1 MPa are the temperature and pressure of the reference state and $H_0$= 10.117 kJ/mol and $S_0$= 261 J/mol/K are the enthalpy and entropy differences between the states at $T_0$ and $P_0$ and at 0 K \cite{wagman1982nbs}. 
$k_B$ is the Boltzmann constant. $c_p$ = 3.5$k_B$ is the heat capacity for diatomic molecules at constant pressure. $\Delta\mu_\mathrm{I}$ is calculated to be $-$0.34 eV at $T=$ 300 K and $P=$ 1 atm.

\pagebreak

\begin{figure}
\includegraphics[width=180mm]{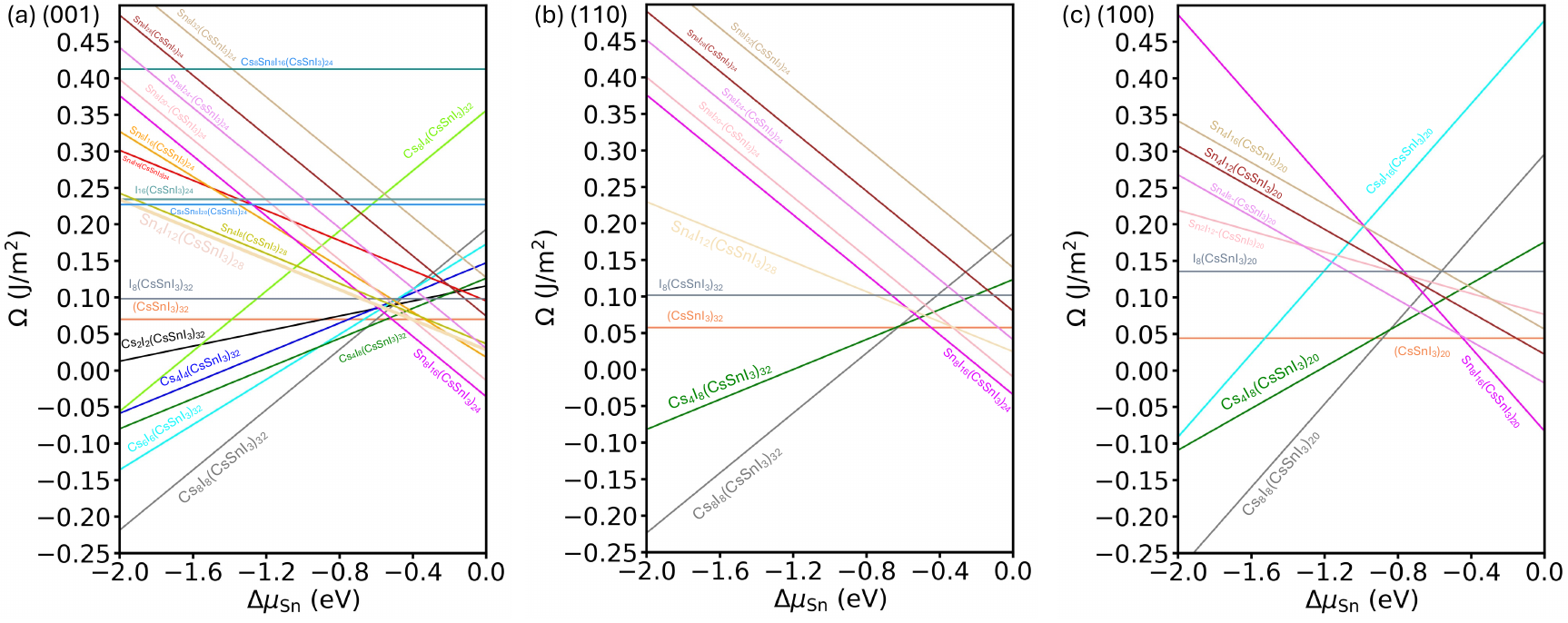}
\caption{\label{fig:eform} The grand potential of each surface is plotted as a function of $\Delta\mu_\mathrm{Sn}$, setting $\Delta\mu_\mathrm{I}=-$0.65 eV for (a) (001), (b) (110), and (c) (100) surfaces of CsSnI$_3$.
}
\end{figure}

\pagebreak

\begin{figure}
\includegraphics[width=170mm]{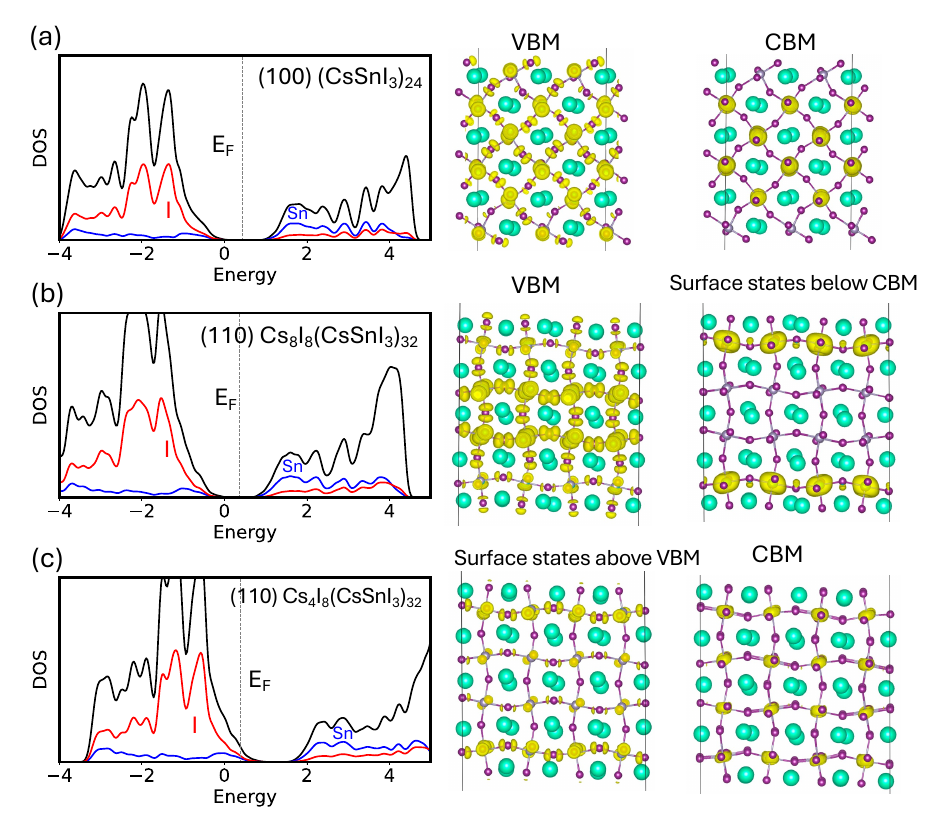}
\caption{\label{fig:pdos} The density of states and partial charge densities of VBM, CBM, or surface states of the (a) (100)-(CsSnI$_3$)$_{24}$, (b) (110)-Cs$_{8}$I$_{8}$(CsSnI$_3$)$_{32}$, and (c) (110)-Cs$_{4}$I$_{8}$(CsSnI$_3$)$_{32}$ surfaces, calculated using PBE+SOC.}
\end{figure}

\clearpage 
\bibliography{supp}{}